\documentclass[sigconf,natbib=true,anonymous=false]{acmart}

\usepackage{times}
\usepackage{latexsym}
\usepackage{booktabs}

\usepackage[T1]{fontenc}

\usepackage[utf8]{inputenc}

\usepackage{microtype}

\usepackage{graphicx}
\usepackage{xcolor}%
\usepackage{multirow}%
\usepackage{amsmath,amsfonts}%
\usepackage{amsthm}%
\usepackage{mathrsfs}%
\usepackage{dsfont}
\usepackage{tikz}
\usepackage{textcomp}
\usepackage{transparent}
\usepackage{pifont}
\usepackage{xspace}

\newcommand{\ses}{\texttt{SEM}\xspace}
\newcommand{\rec}{\texttt{JCRec}\xspace}

\newcommand{\circled}[1]{{\large \textcircled{\small \textbf{#1}}}}

\definecolor{mydarkblue}{RGB}{0,0.08,0.45}

\definecolor{courselight}{RGB}{255,238,238}
\definecolor{joblight}{RGB}{248,255,239}
\definecolor{resumelight}{RGB}{238,249,254}

\usepackage{csquotes}

\usepackage{awesomebox} 
\usepackage{bbding}
\usepackage[most]{tcolorbox}

\usepackage[tikz]{bclogo}
\usepackage[framemethod=tikz]{mdframed}
\definecolor{bgblue}{RGB}{245,243,253}
\definecolor{ttblue}{RGB}{91,194,224}

\newtcolorbox{promptj}{
  colback=joblight!100, 
  colframe=black!40, 
  boxrule=1pt, 
  arc=0pt, 
  boxsep=0pt, 
  left=6pt, 
  right=6pt, 
  top=6pt, 
  bottom=6pt, 
  enhanced, 
  fontupper=\small,
  grow to left by=-1mm,
  grow to right by=-1mm,
  overlay={
    \node[anchor=center, inner sep=0pt, text width=\the\dimexpr\linewidth-12pt\relax, opacity=0.07] at (frame.center) {\includegraphics[width=\linewidth]{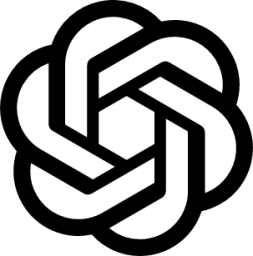}};
  }

}

\newtcolorbox{promptc}{
  colback=courselight!100, 
  colframe=black!40, 
  boxrule=1pt, 
  arc=0pt, 
  boxsep=0pt, 
  left=6pt, 
  right=6pt, 
  top=6pt, 
  bottom=6pt, 
  enhanced, 
  fontupper=\small,
  grow to left by=-1mm,
  grow to right by=-1mm,
  overlay={
    \node[anchor=center, inner sep=0pt, text width=\the\dimexpr\linewidth-12pt\relax, opacity=0.07] at (frame.center) {\includegraphics[width=\linewidth]{figures/openai-logo.png}};
  }
}

\newtcolorbox{promptr}{
  colback=resumelight!100, 
  colframe=black!40, 
  boxrule=1pt, 
  arc=0pt, 
  boxsep=0pt, 
  left=6pt, 
  right=6pt, 
  top=6pt, 
  bottom=6pt, 
  enhanced, 
  fontupper=\small,
  grow to left by=-1mm,
  grow to right by=-1mm,
  overlay={
    \node[anchor=center, inner sep=0pt, text width=\the\dimexpr\linewidth-12pt\relax, opacity=0.07] at (frame.center) {\includegraphics[width=\linewidth]{figures/openai-logo.png}};
  }
}

\AtBeginDocument{%
  \providecommand\BibTeX{{%
    \normalfont B\kern-0.5em{\scshape i\kern-0.25em b}\kern-0.8em\TeX}}}

\copyrightyear{2024} 
\acmYear{2024} 
\setcopyright{acmlicensed}
\acmConference[SIGIR '24]{Proceedings of the 47th International ACM SIGIR Conference on Research and Development in Information Retrieval}{July 14--18, 2024}{Washington, DC, USA}
\acmBooktitle{Proceedings of the 47th International ACM SIGIR Conference on Research and Development in Information Retrieval (SIGIR '24), July 14--18, 2024, Washington, DC, USA}
\acmDOI{10.1145/3626772.3657847}
\acmISBN{979-8-4007-0431-4/24/07}

\begin{document}

\title{Course Recommender Systems Need to Consider the Job Market}

\author{Jibril Frej}
\email{jibril.frej@epfl.ch}
\orcid{0009-0009-0631-0636}
\affiliation{%
  \institution{ML4ED Lab, IC, EPFL}
  \country{Switzerland}
}

\author{Anna Dai}
\email{anna.dai@epfl.ch}
\orcid{0009-0003-7250-1234}
\affiliation{%
  \institution{NLP Lab, IC, EPFL}
  \country{Switzerland}
}

\author{Syrielle Montariol}
\email{syrielle.montariol@epfl.ch}
\orcid{0000-0003-1355-8778}
\affiliation{%
  \institution{NLP Lab, IC, EPFL}
  \country{Switzerland}
}

\author{Antoine Bosselut}
\email{antoine.bosselut@epfl.ch}
\orcid{0000-0001-8968-9649}
\affiliation{%
  \institution{NLP Lab, IC, EPFL}
  \country{Switzerland}
}

\author{Tanja K\"aser}
\email{Tanja.kaser@epfl.ch}
\orcid{0000-0003-0672-0415}
\affiliation{%
  \institution{ML4ED Lab, IC, EPFL}
  \country{Switzerland}
}

\renewcommand{\shortauthors}{Frej, et al.}

\begin{abstract}

Current course recommender systems primarily leverage learner-course interactions, course content, learner preferences, and supplementary course details like instructor, institution, ratings, and reviews, to make their recommendation. 
However, these systems often overlook a critical aspect: the evolving skill demand of the job market. 
This paper focuses on the perspective of academic researchers, working in collaboration with the industry, aiming to develop a course recommender system that incorporates job market skill demands. 
In light of the job market's rapid changes and the current state of research in course recommender systems, we outline essential properties for course recommender systems to address these demands effectively, including explainable, sequential, unsupervised, and aligned with the job market and user's goals.
Our discussion extends to the challenges and research questions this objective entails, including unsupervised skill extraction from job listings, course descriptions, and resumes, as well as predicting recommendations that align with learner objectives and the job market and designing metrics to evaluate this alignment. 
Furthermore, we introduce an initial system that addresses some existing limitations of course recommender systems using large Language Models (LLMs) for skill extraction and Reinforcement Learning (RL) for alignment with the job market. 
We provide empirical results using open-source data to demonstrate its effectiveness.

\end{abstract}




\received{20 February 2007}
\received[revised]{12 March 2009}
\received[accepted]{5 June 2009}

\maketitle

\section{Introduction}


The contemporary job market is dynamic and rapidly evolving~\cite{deming2020earnings}, necessitating continuous adaptation of individual skill sets to maintain relevance and competitiveness. 
This evolution introduces a unique challenge: guiding learners in selecting educational courses that enhance their expertise and align with their career objectives and with job market demands. 
However, a notable mismatch exists between the skills learners possess, the skills taught, and those in demand in the job market~\cite{palmer2017jobs}. This mismatch can be explained by various factors such as the lag between demand in the market and the adaptation from course providers, unequal access to training, or the lack of information from training providers on the type of skills needed in the job market~\cite{palmer2017jobs}.
This issue significantly limits the employability and career progression of individuals, impacting both their personal development and, at a more systemic level, economies dependent on a skilled workforce.


However, existing course recommender systems often focus solely on learner-course dynamics~\cite{SanguinoPerez_Manrique_Mariño_LinaresVásquez_Cardozo_2022, sakboonyarat2019massive}, neglecting the crucial aspect of aligning course recommendations with real-time job market trends. 
This limitation leads to a mismatch between the skills acquired through recommended courses and those in demand in the job market.
Moreover, while some approaches to career path recommendation or skill recommendation consider the job market, users' skills, and goals for their recommendations~\cite{sun2021cost,ghosh2020skill}, they do not recommend specific courses to help users achieve their goals.
Finally, to our knowledge, a single study proposes to use job postings and resumes in a course recommender system~\cite{10.1007/978-3-031-21569-8_7}. However, this approach does not directly consider market trends or users' goals, making this domain largely unexplored.

In this paper, we present the perspective of academic researchers working in collaboration with industry practitioners aiming to develop and deploy job-market-oriented course recommender systems. We argue that rethinking course recommender systems to consider the job market has the potential for significant economic and societal impact. We outline the properties these systems must have: (\textbf{P1}) aligned with the latest job market trends to prioritize courses teaching high-demand skills; (\textbf{P2}) unsupervised to avoid the resource-intensive process of collecting and annotating up-to-date data; (\textbf{P3}) sequential to recommend a sequence of courses where each course builds upon the knowledge acquired in the preceding ones; (\textbf{P4}) aligned with users' goals such as attaining a specific role or increase their marketability; (\textbf{P5}) explainable to ensure user trust and engagement.
We also highlight research directions and areas for future research to address the challenges in this field: (\textbf{RD1})~Addressing the scarcity of course recommendation datasets by creating or providing datasets to the community for training and evaluation; (\textbf{RD2})~Designing evaluation metrics to take into account alignment with the job market when evaluating the recommendations. (\textbf{RD3})~Estimating user's goal progress based on their profile and the job market to tailor the recommendations to their needs; (\textbf{RD4})~Developing Skill-based explainable models and visualization techniques; (\textbf{RD5})~Developing Unsupervised Skill Matching models to estimate up-to-date skill demand on the job market; (\textbf{RD6})~Unsupervised Taxonomy Construction to include new emerging skills without human labeling. 

In this work, we also develop a skill extraction and matching (\ses) method to identify skills and proficiency levels from learners' resumes, course content, and job descriptions. We also develop an unsupervised, sequential skill-based Job-Market-Oriented Course Recommender system (\rec) that uses the skills extracted by \ses to determine a candidate's course options and to estimate the number of job opportunities available to them. \rec then uses Reinforcement Learning (RL) to recommend a sequence of courses that maximizes the number of job opportunities available to the user. Our system meets all of the properties we previously outlined and lays out further steps for the research directions we've mentioned.

The key contributions of our paper include\footnote{Our code is available at \url{https://github.com/Jibril-Frej/JCRec}}:
\begin{itemize}
\item Identification of the desirable properties a job-market-oriented course recommender system should have.
\item Identification of the challenges that developing such systems will pose along with research direction for the community to address these challenges. 
\item A few-shot skill extraction method to find skills from resumes, job postings, and course descriptions.
\item A formulation of job-market-oriented course recommendation as a Markov Decision Process
\item A first job-market-oriented course recommender system.
\end{itemize}

\section{Perspective}


This paper presents the perspective of academic researchers working in collaboration with industry practitioners aiming to develop and deploy course recommender systems for a multilingual user base. This work is motivated by the fact that academic work about course recommender systems seldom considers the job market reality (see section~\ref{sec:courserec}). This disconnection is reflected in the features used to perform the recommendation, but also the lack of consideration of the objective behind the users taking courses. In this work, following extensive collaboration between academic researchers and industries from the up-skilling and continuing education domain, we propose methods to devise recommender systems that align with users' pre-existing skills, career objectives, and the current demands of the job market. Thus, we focus on the challenging and realistic situation of the career development of users, where users have experience and seek to obtain new skills, often to find a new position. 
With this work, we aim to motivate the importance of this problem, propose research directions for the Information Retrieval and Recommendation communities, and provide a prototype of a job-market-oriented course recommender system.

\section{Course Recommender Systems}
\label{sec:courserec}
Course recommender systems have been extensively studied, focusing on various aspects such as learning activities recommendation through open learner models (OLMs)~\cite{DBLP:conf/lak/AbdiKSG20}, recommendation incorporating users' skills~\cite{sankhe2020skill,piao2016analyzing}, peer learner recommendation~\cite{10.1145/3170358.3170400} or target course-oriented recommendation~\cite{10.1145/3303772.3303814}. 
In the recent domain of course recommender systems using neural networks (NN), multiple research directions have been pursued. These include optimizing the accuracy of recommendation~\cite{SanguinoPerez_Manrique_Mariño_LinaresVásquez_Cardozo_2022, 2022.EDM-posters.86}, ensuring fairness~\cite{marras2021, khalid2021novel}, and improving explainability~\cite{10.1007/978-981-19-4453-6_4, 9439852, frej2023finding}. The majority of these approaches use a combination of learner-course interactions, course content, learner preferences, and additional course information such as teachers, schools, course ratings, or comments, usually in the form of Knowledge Graphs (KG). 
While we acknowledge the significance of incorporating learner and course data, we consider that an effective course recommender system must incorporate the job market's current demands, and avoid recommending courses that teach skills lacking demand on the job market.

To our knowledge, \textit{Skill scanner}~\cite{10.1007/978-3-031-21569-8_7} is the only work that uses resumes, course descriptions, and job descriptions for skill-based course recommendations. 
This system's pipeline involves extracting, vectorizing (using word embeggins techniques such as Word2vec~\cite{mikolov2013distributed} and Glove~\cite{pennington2014glove}), clustering, and matching skill sets. \textit{Skill scanner} is used to compare courses, learners, and job postings. 
These comparisons serve not only for skill-based course recommendations but also to inform job seekers and educational institutions about the market relevance of specific skills, enabling them to adapt accordingly. However since their approach relies solely on encoding skill sets in a common representation space, it does not directly consider the job market trends, skill demand distribution, or the user's goal.
Hence, \textit{Skill scanner} is a first step in the direction of job market-guided course recommendations but most of the work remains to be done.

\section{Rethinking Course Recommender Systems to Consider the Job Market}
In this section, we first list the properties that job-market-oriented course recommender systems should have and we propose several Research Directions to address the issues we identified to develop such systems. We voluntarily omitted some properties that all recommender systems should satisfy as they are not specific to our case, such as personalized and real-time recommendations.

\subsection{Properties}
\label{subsec:properties}
\textbf{P1: Job Market Alignment.} The recommender system must consider the skill demand in the job market when making recommendations. On the job market, skills differ by their popularity. Learning a new skill, depending on whether that skill is in high or low demand from employers, or if it is rare or frequent among other job applicants, will have a very different impact on the learner's marketability. Thus, when for example comparing similar courses, the recommender system should give preference to the one teaching the skill with greater market demand. This ensures that learners are equipped with the most relevant and sought-after skills to increase their chances of finding a position.\vspace{0.5em}\\
\textbf{P2: Minimal Supervision.} The recommender system should rely on a limited amount of labeled data because it needs to accommodate the rapid evolution of the job market with new skills appearing regularly. However, existing recommenders usually rely on supervised models that would have to be updated regularly. To avoid the high cost of labeling data manually, most of the components should be based on unsupervised learning techniques. For example, scraping job postings and extracting skills from course descriptions using unsupervised models will allow the system to adapt to market trends without the extensive costs of manual data labeling.\vspace{0.5em}\\
\textbf{P3: Sequential Recommendations.} The system must recommend sequences of courses rather than standalone courses. Indeed, individual goals often require a progression through multiple subjects. Note that the order in which these courses are taken often matters.
For instance, a front-end developer aspiring to become an LLM engineer might need courses in Python, machine learning, Natural Language Processing (NLP), and LLMs, with each course building upon the knowledge acquired in the preceding ones.\vspace{0.5em}\\
\textbf{P4: User's Goal Alignment.} The system must align with the objectives of its users, whether that involves attaining a specific role, learning skills to increase overall profile attractiveness, or specializing in a field. Recommendations should thus be aligned with these goals, ensuring that two users with identical profiles but different objectives receive different course suggestions.\vspace{0.5em}\\
\textbf{P5: Explainable.} The recommender system must provide explainable recommendations. Given the time and resources needed for a user to enroll and finish a course, it is crucial for the system to transparently explain its recommendations -- especially sequential recommendations. Explainability is not only essential for building trust but also for ensuring that users feel confident in their decision to invest resources in a recommended course. For example, explaining that a course is suggested because it teaches a skill with current high demand in the job market can significantly enhance user confidence in the system's guidance.

\subsection{Research Directions and Challenges}

\textbf{RD1: Course Recommendation Datasets}
Beyond consideration of the job market, a significant challenge affecting research in course recommendation is the scarcity of publicly accessible, large-scale datasets for this task. Presently, there are only two of such datasets: \texttt{Xuetang}~\cite{DBLP:conf/aaai/ZhangHCL0S19} and \texttt{COCO}~\cite{cocodata}. \texttt{Xuetang} contains courses from the Massive Open Online Courses (MOOCs) platform XuetangX\footnote{\url{https://next.xuetangx.com/}}, primarily in Chinese. \texttt{COCO} contains courses from the MOOCs platform Udemy\footnote{\url{https://www.udemy.com/}} available in 35 languages. This sparsity can be attributed to universities' reluctance to share student enrollment data due to privacy and ethical concerns and the desire of online course providers to preserve competitive advantages.
In response to these challenges, we urge academics and industry professionals to anonymize and share subsets of their enrollment data with the research community. An alternative method we wish to highlight is the generation of synthetic datasets using generative models. While this method has been proposed in the context of job postings for skill extraction using LLMs~\cite{decorte2023extreme,clavie2023large,magron2024jobskape}, its application to creating course recommendation datasets remains unexplored. This presents an innovative research direction with the potential to significantly impact the field. Potential directions could involve generating coherent career paths of individuals using the in-context ability of LLMs, then inferring skills required to switch from one position to the next one using databases of skills associated with job roles.\footnote{A potential data source for this in the IT domain is \url{https://www.berufe-der-ict.ch/berufe-der-ict/}.}
Moreover in this work, instead of course enrollment data, we use only course descriptions along with job postings and resumes. While job postings are regularly explored in the literature~\cite{zhang2022skillspan,zhang-etal-2022-kompetencer,magron2024jobskape}, few public datasets exist. User profiles, in the form of resumes, are documents that contain personal information and are usually not public; such databases are even harder to come by than job postings, especially at scale. Overall, these data sources are seldom found in languages other than English. We show that we can use such data to build job-market-guided recommender systems, even though it doesn't allow for direct evaluation of users' course enrollment.\vspace{0.5em}\\
\textbf{RD2: Evaluation}
Evaluating the effectiveness of a job-market-oriented course recommendation poses significant challenges. Most recommender systems estimate the relevance of a course based on user profiles using ranking metrics (MRR, nDGG, Hit) to compare recommendations to the actual courses taken using a user-course interaction matrix. Other metrics are used to measure Novelty, Fairness, Diversity, and Coverage of the recommendations~\cite{zangerle2022evaluating}. However, these approaches are not enough for job-market-oriented course recommendations as they do not take into account market demand and the actual impact on a user's career trajectory. An ideal evaluation framework would also ascertain whether the recommended course enabled the user to meet their career goals, such as securing a desired job or enhancing their marketability. Because this type of information is rarely directly available, estimating the impact of following a course on a user's profile and achieving their objective presents a considerable challenge. For these reasons, we believe that designing an evaluation methodology for job-based course recommender systems is both a challenging and impactful research problem. In this work, we propose to evaluate the recommendations by estimating the number of jobs the users can apply to upon completing the recommended courses. Although our approach presents a first step for this research direction, user studies are necessary to evaluate if the system provides meaningful recommendations for users. Another potential research direction involves addressing the following problem: how can historical user enrollment data be leveraged to assess the effectiveness of a recommender system in a dynamic job market where past user choices may become less relevant? For instance, consider a scenario where a user previously enrolled in a \texttt{PHP} course, leading to employment in web development. Recommending this same course in the present day may not hold the same relevance, given shifts in the job market. Using this interaction for evaluation might bias the system toward past trends but ignoring all past interactions is not a satisfying solution since course enrollments include valuable information on user preferences that provides insight into specific flaws of evaluated recommender systems. One possible direction would be to ignore the skills of the courses -- which might teach outdated skills-- and focus on the user enrollment data, evaluating only the coherence and robustness of recommender systems between users.\vspace{0.5em}\\
\textbf{RD3: Estimating User's Goal Progress}
On top of incorporating the dynamics of the job market in the recommendation, we highlight the importance of taking into account the objectives of learners.  We propose an initial strategy involving two steps: 1) identifying the goals learners may hold, and 2) developing functions to assess how much progress learners have made towards achieving their goals.
To tackle this, we identify three primary goals related to the job market that a user might pursue as well as the functions needed to estimate their progress towards their goal:\\
\textbf{1)} Securing a Specific Position: we need to define a function that can compute the compatibility between a learner's profile and a job description.\\
\textbf{2)} Boosting Marketability: we need to define a function to evaluate a learner's marketability using their profile and a collection of job descriptions.\\ 
\textbf{3)} Achieving Specialization in a Field: We also need a function to quantify a learner's level of specialization based on their profile and the target field.\\
Defining these functions, along with the constraints and properties they must satisfy~\cite{DBLP:conf/sigir/FangTZ04}, represents a promising and impactful direction for research. This approach aims to align course recommendations more closely with the realities of the job market and the individual goals of learners, thereby enhancing the relevance of such systems.\vspace{0.5em}\\
\textbf{RD4: Skill-based Explainability}
Given the significant time and investment required to enroll in a course, such systems must offer clear explanations. Users need to understand how enrolling will advance them toward their goals. We advocate for a skill-based approach to explainability, highlighting the specific skills a course teaches and how acquiring these skills aligns with the user's objectives.
Possible research directions include: 1) exploring visualization techniques to present the skill-based explanations to the learners and 2) developing sequential and explainable course recommender systems. To our knowledge, recommender systems that are both sequential and explainable have not been studied, presenting a novel research opportunity for the field of recommender systems.\vspace{0.5em}\\
\textbf{RD5: Unsupervised Skill Matching}
Skill matching consists of aligning skills from various sources, such as job descriptions, resumes, and courses, with a skill taxonomy~\cite{gnehm-etal-2022-fine}. Because this process is crucial for creating skill-based explanations in job-oriented course recommender systems, we highlight the necessity of unsupervised methods to extract skills from resumes, course descriptions, and job postings. Recent advancements in LLMs have facilitated few-shots skill extraction from job postings through the use of synthetic dataset generation to generate demonstrations~\cite{magron2024jobskape}. However, extending this approach to include resumes and course descriptions remains largely unexplored. Possible research directions encompass generating synthetic datasets for resumes and course descriptions to enable unsupervised skill matching. Another direction is to assess skill proficiency levels from resumes, job postings, and course descriptions. Such assessments could give the recommender system's ability to distinguish between introductory and advanced courses. To our knowledge only one supervised model has been used for estimating skill levels from resumes~\cite{banerjee2022estimating}, leaving a gap in research for unsupervised approaches and other types of sources such as job postings and course descriptions. This area offers significant potential for impactful research, given its unexplored status.\vspace{0.5em}\\
\textbf{RD6: Unsupervised taxonomy construction}
A skill-based recommender system that would flexibly adapt to evolutions of the job market requires, on top of an unsupervised skill matching tool, a method to adapt the skill taxonomy to emerging market trends. Constructing and maintaining a skill taxonomy manually is a time-consuming and labor-intensive task. Moreover, it requires state-of-the-art expertise in the domain of emerging skills. To address this challenge, unsupervised methods for automatically constructing and updating skill taxonomies can be implemented. In a practical setting, this would be performed during the skill matching step, allowing the identification of skills that fail to be matched with taxonomy items and automatically adding them to the taxonomy \cite{xu2023tacoprompt, takeoka2021low}.

\section{Job-Oriented Course Recommender}
In this section, we describe a first system that satisfies some of the properties we presented in section~\ref{subsec:properties}.
Our methodology comprises two steps: Unsupervised Skill Extraction and Matching (\ses), followed by Job-Market-Oriented Course Recommender (\rec). The key question addressed by these experiments is whether, for a small company, the investment of resources and time in using RL to develop a job-oriented course recommender system is justified, or if a Greedy heuristic would suffice.

\begin{figure*}
    \centering
    \includegraphics[width=\textwidth]{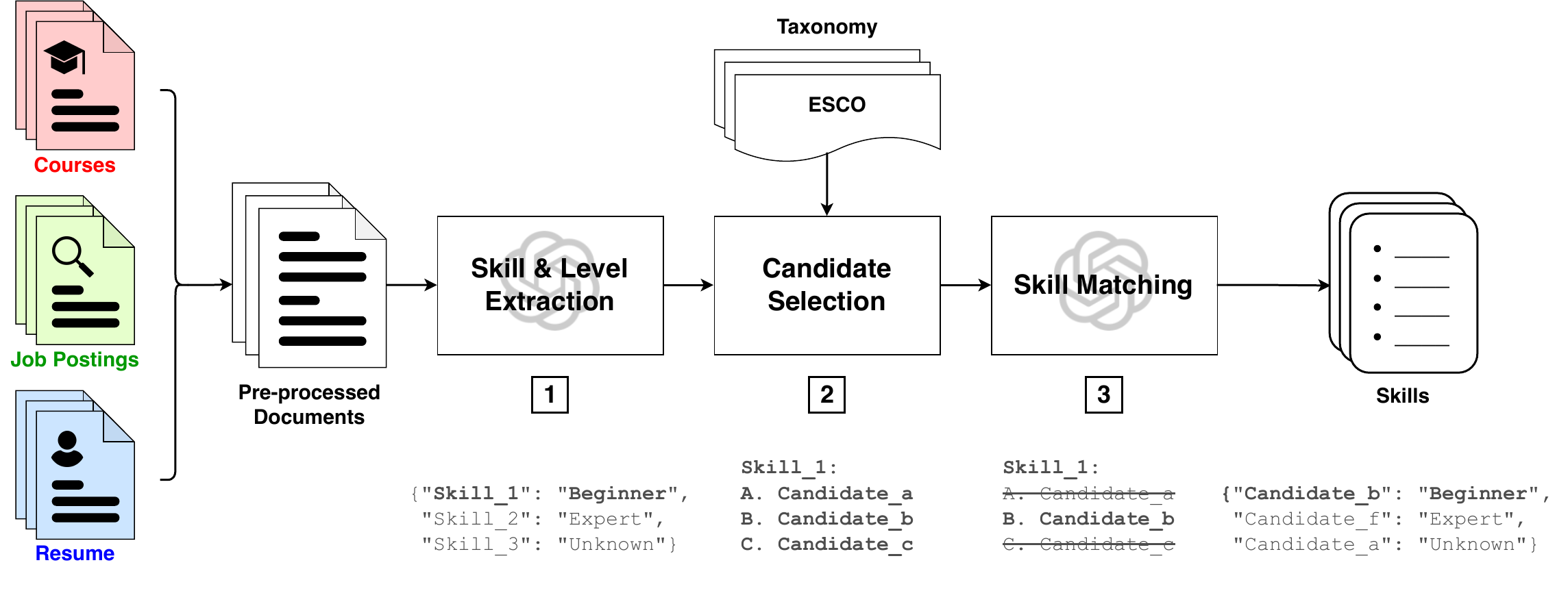}
    \caption{Illustration of the skill extraction and matching (\ses) pipeline. \fbox{\textbf{1}} Given a document, \ses\ extracts relevant skills and proficiency levels from using LLM prompting. \fbox{\textbf{2}} For each extracted skill, three candidates from ESCO taxonomy are selected using string matching and embedding similarities. \fbox{\textbf{3}} The extracted skill and taxonomy candidates are prompted to an LLM to find the best match.}
    \label{fig:Extraction}
\end{figure*}

\subsection{Skill Extraction and Matching}

Following \cite{magron2024jobskape}, we use an LLM-based system to extract skills from documents (job postings, course descriptions and resumes) and match these skills to the ESCO taxonomy~\cite{le2014esco}. We augment this pipeline with a proficiency levels extraction step, classifying each extracted skill into four categories: ``beginner", ``intermediate", ``expert" and ``unknown''. Because \ses is unsupervised, works on job postings, resumes, and course descriptions and estimates the proficiency level of skills, it satisfies property \textbf{P3} while providing a first step for research direction \textbf{RD5}.

As illustrated in Figure~\ref{fig:Extraction}, the \ses\ pipeline, follows three steps:\\
\fbox{\textbf{1}}~\textbf{Skill \& level Extraction} from the document;\\
\fbox{\textbf{2}}~\textbf{Candidate Selection} from the taxonomy;\\
\fbox{\textbf{3}}~\textbf{Skill Matching} from the candidates;\\ 
This multi-step approach allows us to flexibly handle a taxonomy of any size without facing context window limitations. In the following, we describe in depth each of these steps.\vspace{0.5em}\\
\fbox{\textbf{1}}~\textbf{Skill \& level Extraction.} We first break down each document into individual sentences that are then grouped into sets of one or two, which are processed through the pipeline sequentially. This approach provides the LLM with some context while addressing constraints related to document length.
Next, we utilize the LLM in a six-shot setting to identify skills and their associated proficiency levels within these sentence groups.
To ensure an effective extraction, our few-shot demonstrations are deliberately varied. They include a range of examples: positive and negative instances, answers that are both multiple and single, as well as examples of both hard and soft skills drawn from various contexts~\cite{nguyen2024rethinking}.\vspace{0.5em}\\
\fbox{\textbf{2}}~\textbf{Candidate Selection.}
We select skill candidates from the ESCO taxonomy based on their names and definitions using two methods:\\
1) rule-based: We assume that if the extracted skill appears exactly in the taxonomy, it will be a good candidate for a match. When an exact match is not found, we use the \texttt{token\_set\_ratio} method from TheFuzz\footnote{https://github.com/seatgeek/thefuzz}, which computes the similarity by comparing shared and unique tokens between the extracted and taxonomy skills.\\
1) embedding-based: We compute the cosine similarity between the extracted skill and the taxonomy skills using \texttt{JobBERT}~\cite{decorte2021jobbert}.

The candidates are selected by taking the union of the top three ESCO skills returned by each method. 
We chose a hybrid approach because of the complementary merits and limitations of the two methods.
The rule-based method selected viable candidates but missed any synonyms, whereas the embedding-based method risked selecting contextually similar yet factually dissimilar candidates.\vspace{0.5em}\\
\fbox{\textbf{3}}~\textbf{Skill Matching.}
Extracted skills are matched to the selected candidate's skills. Formatted candidate options (e.g., A. B. C.) are presented to the LLM, which identifies the best match if any, or indicates no match.

\subsection{Recommendation}
\label{subsec:recommendation}

\begin{figure*}
    \centering
    \includegraphics[width=\textwidth]{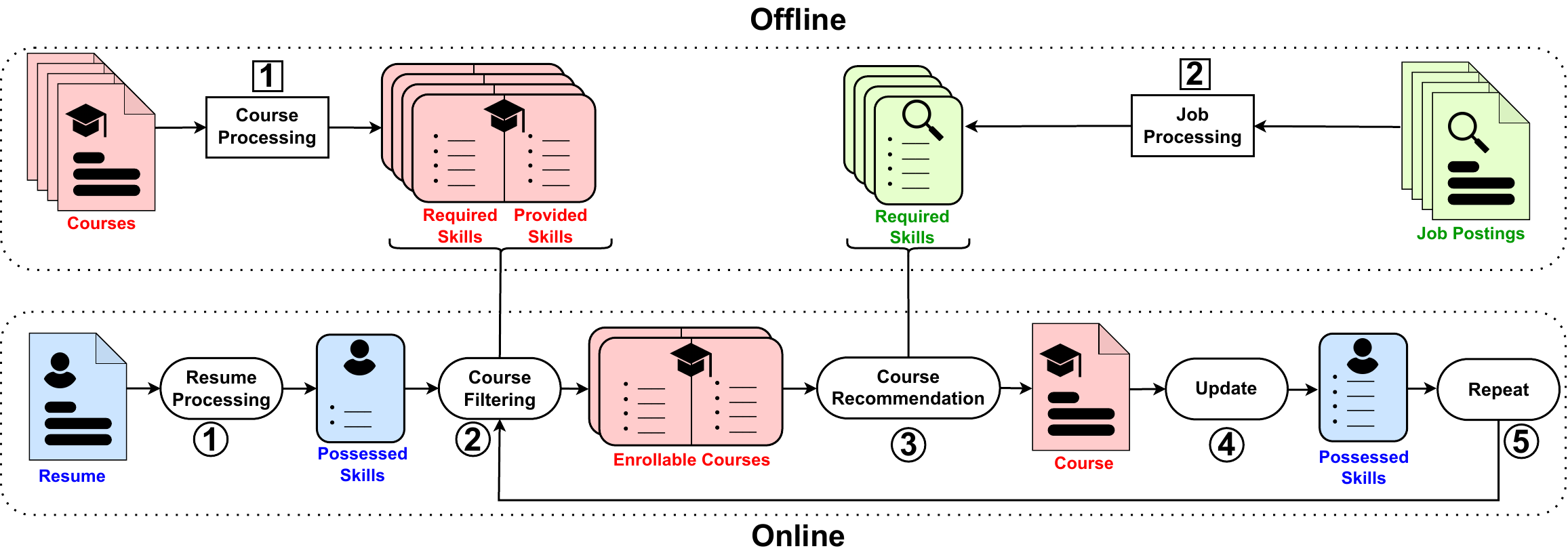}
    \caption{Illustration of the \rec pipeline: In the offline phase, \ses\ is used to \fbox{\textbf{1}} extract skills required to take each course and skills provided by the courses, and to \fbox{\textbf{2}} extract skills required by each job posting skills. During the online phase, as a user uploads their CV, \circled{1} \ses\ extract their skills. \circled{2} These skills are used to filter the set of courses they can enroll in. \circled{3} From these, one course is recommended, aiming to maximize the increase in the number of jobs the user can apply for. \circled{4} The user's profile is then updated with the skills acquired from the recommended course. \circled{5} In the case of sequential recommendation, steps \circled{2}, \circled{3}, and \circled{4} are repeated until $k$ courses are recommended.}
    \label{fig:RecPipeline}
\end{figure*}

As illustrated in Figure~\ref{fig:RecPipeline}, \rec has two main components: an offline preprocessing phase that uses \ses, and the online recommendation. Our system is fully unsupervised, makes explainable skill-based sequential recommendations, and assumes that the user's goal is to maximize their marketability. These characteristics make our system satisfy properties \textbf{P1}, \textbf{P2}, \textbf{P3}, \textbf{P5}, and partially \textbf{P4} since we assume a specific goal for the user. \rec is also an additional step for research directions \textbf{RD3} and \textbf{RD4}, although user studies will be required to assess the explainability of our approach in practice. In the following, we describe in depth \rec.\vspace{0.5em}\\
\textbf{\large{Offline Preprocessing}}
The preprocessing phase involves extracting skills and proficiency levels from courses and jobs.\vspace{0.25em}\\
\fbox{\textbf{1}}~\textbf{Course Processing.}
In this phase, we use \ses\ on each course. This method identifies all the skills required for enrollment and the skills that will be acquired upon completion. As a result, we obtain a set of processed courses $\mathcal{C}$. Each course $c \in \mathcal{C}$ is represented by two sets: $c_r$ the required skills with their proficiency levels, and $c_p$ the provided skills with their proficiency levels. We associate each proficiency level with a positive integer. For instance, the course \textit{Mastering Python: from beginner to expert} might be represented as $(c_r:\{(python, 1)\}; c_p:\{(python, 3)\})$, with the beginner level corresponding to integer 1 and the expert level to integer 3.\vspace{0.25em}\\
\fbox{\textbf{2}}~\textbf{Job Processing.}
A similar preprocessing is applied to jobs, resulting in a set of processed jobs: $\mathcal{J}$. Each job $j \in \mathcal{J}$ is associated with a set of required skills and proficiency levels. For example, a \textit{Data Engineer} job posting could be represented as $j:\{(python, 2), (SQL, 1)\}$.\vspace{0.5em}\\
\textbf{\large{Online Recommendation}}
Using the processed courses and jobs, we can proceed to the online recommendation.\\
\circled{1}~\textbf{Resume Processing.}
\rec begins by processing the user's resume. Using \ses, it extracts the user's skills and proficiency levels, creating a set $u$ consisting of pairs of skills and their corresponding proficiency levels. For instance, a junior data scientist's skill set might be represented as $u:\{(python, 2), (machine learning, 2)\}$.\\
\circled{2}~\textbf{Course Filtering.}
In the second step, we use a user-course relevance function $\text{ucr}(u, c)$ to determine $\mathcal{C}_u$: the set of courses available for enrollment by user $u$. The relevance function $\text{ucr}(u, c)$ is defined in Appendix~\ref{subsec:appendix:lcsim}, along with the desirable properties it should satisfy. A threshold $t_{uc}$ is set to filter courses, keeping those with a relevance score higher than $t_{uc}$. Thus, $\mathcal{C}_u$ is defined as $\mathcal{C}_u  = \{c \in \mathcal{C} |\; \text{ucr}(u, c) \geq t_{uc}\}$.\\
\circled{3}~\textbf{Course Recommendation.}
Next, we recommend a course $c \in \mathcal{C}_u$. Given the nature of our problem as a sequential decision-making task, we use Reinforcement Learning (RL), as commonly done for sequential recommender systems~\cite{afsar2022reinforcement}. We define our problem's Markov Decision Process (MDP) as follows: the state is $u$ the user's skill set; the action space is $\mathcal{C}$, the set of courses; and the reward to maximize is the marketability that we define as the number of jobs the user can apply to, denoted $|\mathcal{J}_u|$. Moreover, if the agent recommends a course that is not in $\mathcal{C}_u$, i.e. that the user cannot or should not follow, the reward is set to -1 and the episode is terminated. This choice of reward is motivated by the fact that we are only considering users whose goal is to increase their marketability. If different goals were considered, additional rewards would be required. Finally, the transition probabilities from state to state are deterministic: completing a course adds skills $c_p$ to $u$, resulting in probabilities of either 0 or 1.\\
\noindent\circled{4}~\textbf{Update} After recommending course $c$, the user's skills $u$ are updated with $c_p$: the skills provided by course $c$. The number of jobs user $u$ can apply to $|\mathcal{J}_u|$ is computed using the similarity function $\text{ujs}(u, j)$ defined in Appendix~\ref{subsec:appendix:ljsim}. To determine $|\mathcal{J}_u|$, we set a threshold $t_{uj}$ and consider that a user $u$ can apply to a job $j$ if and only if $\text{ujs}(u, j) \geq t_{uj}$, making the set of applicable jobs defined as: $\mathcal{J}_u  = \{j \in \mathcal{J} |\; \text{ujs}(u, j) \geq t_{uj}\}$. \\
\noindent\circled{5}~\textbf{Repeat} To recommend a sequence of courses, we repeat steps \circled{2}, \circled{3}, and \circled{4} until $k$ courses have been recommended.

\section{Experimental Setup}

\subsection{Datasets}
\label{sec:data}
In our experiments and evaluation, we used publicly available datasets. We concentrated specifically on English-language documents within the information technology (IT) and IT management industry, due to the ease of access and relevance to our study's focus.\vspace{0.25em}\\
\textbf{Taxonomy.} We use the ESCO~\cite{DBLP:ESCO} taxonomy that comprises 13\,890 skills and competencies. We filter out competencies irrelevant to IT and IT management roles, resulting in a subset of 1\,794 skills.\vspace{0.25em}\\
\textbf{Job Postings.} We scraped around 3,500 English job postings in the domain of Information Technology from various online platforms.\vspace{0.25em}\\
\textbf{Course Descriptions.} The \texttt{COCO} dataset~\cite{cocodata} contains 43\,113 online Udemy courses, with their descriptions, objectives, and prerequisite requirements. The dataset is divided into 133 detailed categories and covers 46 languages. We first filter the dataset for only English courses (34\,111 courses). Then, we filter the granular categories to include only the 27 that are related to IT and IT management (leaving 12\,291 courses). In our experiments, we took a random sub-sample of 3\,000 courses.\vspace{0.25em}\\
\textbf{Resumes.} We used two resume datasets from Kaggle for this study. The first dataset\footnote{\url{https://www.kaggle.com/datasets/snehaanbhawal/resume-dataset}} comprises 2\,484 resumes spanning 24 industries. We filter the dataset to include only the \textit{Information-Technology} category. The second dataset\footnote{\url{https://www.kaggle.com/datasets/gauravduttakiit/resume-dataset.}} contains 166 resumes of professionals working in the IT department. We exclude resumes from non-IT personnel (i.e., in \textit{HR}, \textit{Arts}, or \textit{Health and fitness} categories). Together, we keep a subset of 233 anonymized resumes from IT professionals.

\subsection{Pre-processing}

\begin{table}
    \centering
    \footnotesize
    \begin{tabular}{lcccc}
        \toprule
        \multirow{2}{*}{} & {\textbf{Jobs}} & \multicolumn{2}{c}{\textbf{Courses}} & {\textbf{CVs}} \\
        \cmidrule(lr){3-4}
        \textit{average} & & Prereqs & Target & \\
        \midrule
        Words per sentence & 19.4 & 19.5 & 32.7 & 23.7 \\
        Sentences per doc & 26.9 & 2.0 & 2.4 & 29.7 \\
        Words per doc & 510.5 & 38.6 & 77.8 & 702.4 \\ 
        \bottomrule
    \end{tabular}
    \caption{Statistics of post-processed documents}
    \label{tab:sentence_stats}
\end{table}

For job postings and course descriptions, we remove the document if the combined text body contains less than 50 and 20 words respectively, a cutoff driven by examinations of documents below these thresholds. 
For courses, we split descriptive textual features into two categories: text describing prerequisite skills to take the course (``requirements'' field), and text describing skills taught in the course (``course objectives'' and ``course description'').
In addition, we chose to include only short descriptions and not long descriptions of the \texttt{COCO} courses because we found too many irrelevant and repetitive keywords mentioned in the long descriptions.
With each document type processed, we segment the text bodies into individual sentences for analysis by \ses\. Table~\ref{tab:sentence_stats} shows the varying lengths of these documents, indicating notably longer job postings and resumes compared to courses. This difference arises as we only use short course descriptions.
Furthermore, through qualitative analysis, we find that sentences in resumes tend to be independent while those in jobs and courses are contextual. Hence, we process each resume sentence independently, while job and course sentences are handled in pairs.

\subsection{Skill matching setup}

We run the job postings, course descriptions, and resumes through the matching pipeline, from which we extract 165K, 30K, and 31K total skills respectively. While we do not have annotated data, we can examine the results with heuristics. Table~\ref{tab:skills_stats} indicates that resumes contain the most extracted skills but show the lowest percentage of proficiency level identification per skill. This trend aligns with common resume practices in the IT sector, where professionals often list numerous skills without specifying their proficiency levels. Moreover, course prerequisites and descriptions, as shown in Table~\ref{tab:skills_stats}, feature the fewest skills. This aligns with the expectation that courses have a limited scope, making the teaching of a vast number of skills impractical. Conversely, the number of skills extracted from job postings at 43.2 skills per document, or about 1.6 skills per sentence is justifiable. IT sector job postings often list many required skills, reflecting this sector's diverse skill demands.

\begin{table}
    \centering
    \begin{tabular}{ccccccc}
        \toprule
        \multirow{2}{*}{} & {\textbf{Jobs}} & \multicolumn{2}{c}{\textbf{Courses}} & {\textbf{CVs}} \\
        \cmidrule(lr){3-4}
        \textit{average per doc} & & Prereqs & Target & \\
        \midrule
        Skills extracted & 43.2 & 3.2 & 6.8 & 112.9 \\
        Levels extracted & 71.1\% & 72.0\% & 79.6\% & 45.4\% \\
        Skills matched & 16.6 & 2.2 & 4.2 & 22.6 \\
        \bottomrule
    \end{tabular}
    \caption{Statistics on skills standardized by document type: \textit{"Skills extracted"} shows the average unique skills extracted, \textit{"Levels extracted"} indicates the average percentage of skills with identified levels, and \textit{"Skills matched"} represents the average unique skills matched to the taxonomy. }
    \label{tab:skills_stats}
\end{table}

We also validated the performance of the proficiency level extraction method. An expert manually annotated 212 skills from 10 job postings in the IT sector. We compared the annotations with the levels extracted by our system and obtained an accuracy of 0.68.

Furthermore, we explore the results of the extraction on each type of document. Table~\ref{tab:level_stats} shows the percentage of the extracted proficiency levels that are "expert", "intermediate", "beginner", "unknown", or "other". Levels found in job descriptions are primarily "expert" indicating when employers suggest desired skill levels, nearly 50\% of the time, they look for experts in that skill. This is expected and agrees with qualitative observations of job descriptions. More than 54\% of levels from resumes are "unknown", which is reasonable as stated previously. However, when IT professionals do indicate their skill levels, it is also expected that skills are either at the "expert" level or "intermediate" level. This also agrees with qualitative observations, since professionals tend not to list a skill unless they are proficient. Looking at courses, levels of skills found in both course prerequisites and objectives are primarily "beginners". This is expected because online courses tend not to be in-depth and are geared toward beginners. 
It is also expected that very few courses (less than 10\%) have intermediate or expert level prerequisites, whereas over 30\% of courses teach intermediate or expert level skills. 
Finally, we observe that despite indicating in the prompt to output only specific words, all documents output a small percentage (under 1\%) of non-conforming levels.

\begin{table}
    \centering
    \begin{tabular}{lcccc}
        \toprule
        \multirow{2}{*}{} & {\textbf{Jobs}} & \multicolumn{2}{c}{\textbf{Courses}} & {\textbf{CVs}} \\
        \cmidrule(lr){3-4}
        \textit{\% of all skills} & & Prereqs & Target & \\
        \midrule
        "expert" & \textbf{49.2} & 0.5 & 3.0 & 21.8 \\
        "intermediate" & 19.8 & 9.0 & 29.5 & 20.6 \\
        "beginner" & 2.1 & \textbf{51.7} & \textbf{47.1} & 3.0 \\
        "unknown" & 28.3 & 38.0 & 20.1 & \textbf{54.4} \\
        other & 0.6 & 0.8 & 0.3 & 0.2 \\
        \bottomrule
    \end{tabular}
    \caption{Distribution of extracted skill mastery levels for each document type. other indicate cases where the LLM failed to output a level in one of the 4 predefined categories.}
    \label{tab:level_stats}
\end{table}

Finally, we replace unknown levels with the following heuristic: if a level is unknown for a learner it will be set to beginner and if a level is unknown for a job, it will be set to expert. This heuristic allows us to have a conservative system that will not assume that a beginner would be able to apply to a job that requires expert knowledge. Regarding unknown proficiency levels for courses, we replace them with intermediate and correct the following types of inconsistencies: if a course requires skill $s$ at level $l_r$ and also teaches the same skill $s$ at a lower level $l_p$ such that $l_r \geq l_p$ this implies an inconsistency: the course demands a higher skill level than it provides. In such cases, we adjust $l_r$ to the level before $l_p$ to ensure the removal of any such inconsistencies.
For skill extraction and skill matching, we use \texttt{GPT-3.5-turbo} with temperature set to 0, top-p set to 1.0, frequency penalty set to 0.0, and presence penalty set to 0.0.

\subsection{Recommendation Algorithms}

\begin{table*}
\centering
\begin{tabular}{c|c|cc|cc|cc|cc|cc}
\toprule
\multirow{2}{*}{Model} & $k=0$ & \multicolumn{2}{c}{$k=1$} & \multicolumn{2}{c}{$k=2$} & \multicolumn{2}{c}{$k=3$}  & \multicolumn{2}{c}{$k=4$}  & \multicolumn{2}{c}{$k=5$} \\
\cline{2-12}
 &  \texttt{Rwd} &   \texttt{Rwd} & \texttt{Time} (ms) & \texttt{Rwd} & \texttt{Time} (ms) & \texttt{Rwd} & \texttt{Time} (ms) & \texttt{Rwd} & \texttt{Time} (ms) & \texttt{Rwd} & \texttt{Time} (ms) \\
\hline
Exhaustive & 0.1  & 1.0 & 16 & 2.5 & $10^2$ & 5.6 & $10^4$&NA&NA&NA&NA\\
\hline
Greedy & 0.1 & \textbf{1.0} & 16 & \textbf{2.0} & 41 & \textbf{4.1} & 73 & 7.0 & $10^2$ &\textbf{10.5} & $10^2$ \\
DQN & 0.1 & 0.9 & 4.8 & 1.7 & 8.7 & \textbf{4.1} & \underline{11} & 6.3 & \underline{17} & 7.1 & 24\\
PPO & 0.1 & 0.5 & \underline{3.7} & 1.1 & \underline{6.4} & 3.4 & \underline{11} &  \textbf{8.0} & 20 & 10.4 & \underline{21}\\
\bottomrule
\end{tabular}
\caption{Evaluation of the 4 course recommendation algorithms. \texttt{Rwd} is the reward i.e. the average number of jobs learners can apply to after following the course recommendations. The highest rewards among Greedy, DQN, and PPO are highlighted in bold for each $k$. \texttt{Time} is the average time in milliseconds needed for the recommendations. The lowest time among Greedy, DQN, and PPO is underlined for each $k$. The case $k=0$ reflects the initial learner state, identical across all algorithms. The exhaustive algorithm was not run for $k=4$ and $k=5$ due to estimated times exceeding $10^6$ seconds per learner.}
\label{tab:rec_res_lvl_3}
\end{table*}

In this work, we compare four sequential recommendation algorithms: exhaustive, greedy, and two RL-based algorithms.\\
\textbf{Exhaustive Recommendation}:
The exhaustive approach evaluates all possible course sequences that can be recommended to a learner.  For every possible sequence, it computes $|\mathcal{J}_u|$ the number of jobs the learner can apply to after updating their profile. The sequence that maximizes $|\mathcal{J}_u|$ is then recommended. Although this approach guarantees the recommendation of the optimal sequence, its practical feasibility is limited due to the high time complexity that increases exponentially with the number of recommended courses: $O(|\mathcal{J}|\cdot |\mathcal{C}|^k)$. Nevertheless, we have chosen to present the outcomes of this methodology, despite its impracticality for real-world application, as it provides a theoretical maximum for the value of $|\mathcal{J}_u|$.\\
\textbf{Greedy Recommendation}:
At each of the $k$ steps, the greedy algorithm recommends a course that maximizes the number of job opportunities available to the learner. This approach is greedy because it selects the best immediate option at each step without guaranteeing an overall optimal sequence. Despite this sub-optimality, it is more time-efficient than the exhaustive approach. For each course recommendation, the learner's profile is updated and the similarity with every job is computed to estimate the course leading to the highest increase in job applicability. This results in a total number of operations on the order of $O(k \cdot |\mathcal{J}|\cdot |\mathcal{C}|)$.\\
\textbf{Reinforcement Learning Recommendation}:
We compare 2 RL algorithms: Deep Q Network (DQN)~\cite{mnih2013playing} and Proximal Policy Optimization (PPO)~\cite{schulman2017proximal}. For both RL algorithms, we used the implementation provided by Stable-Baselines3~\cite{raffin2021stable} with default parameter values. The input size is the number of skills $|\mathcal{S}|$ and the output dimension is the number of actions, or courses, $|\mathcal{C}|$. This makes the time complexity of these algorithms $O(k \cdot |\mathcal{S}| \cdot|\mathcal{C}|)$, meaning that RL has the potential to be more efficient than the greedy heuristic, especially if the number of skills $|\mathcal{S}|$ is one or more order of magnitude smaller than the number of jobs $|\mathcal{J}|$.\\
\textbf{Hyperparameters}:
Both threshold values $t_{uc}$ and $t_{uj}$ were set to 0.8, the RL agents were trained for 5\;000\;000 steps.

\subsection{Evaluation}
We evaluate the algorithms using different values of $k$: the number of courses recommended in the sequence. Our evaluation employs two metrics to compare the algorithms:
\textbf{1)} The average number of jobs a learner can apply to after receiving $k$ course recommendations assessing the relevance of the courses recommended, and \textbf{2)} the average time taken to make a recommendation, evaluating the algorithms' feasibility for real-world deployment.
Evaluation is conducted on a subset of the dataset described in section~\ref{sec:data}, including 100 jobs, 100 courses, and all resumes containing less than 15 skills for a total of 52 resumes. We decided to limit the number of skills present in the resume to emulate a learner new to the job market.

\section{Results}

Table~\ref{tab:rec_res_lvl_3} presents a comparison of the Exhaustive, Greedy, DQN, and PPO algorithms across different sequence lengths ($k$). \texttt{Rwd} is the average number of jobs learners can apply to after taking the recommended courses and \texttt{Time} is the average duration for each recommendation in milliseconds.

The Exhaustive algorithm, although delivering the best recommendation quality in terms of reward is not suited for real-world deployment due to its significant time consumption. Even on a small dataset of 100 courses and jobs, it can take up to 10 seconds to recommend a sequence of 3 courses. Its usage becomes unfeasible as the sequence length increases, rendering it unsuitable for practical applications. However, the exhaustive algorithm provides an upper bound on the reward as it outputs the optimal course sequence. Nevertheless, in the remainder of our analysis, we will not compare the exhaustive approach to the greedy and RL algorithms.

For shorter sequences ($k=1, 2, 3$), the Greedy algorithm achieves the highest \texttt{Rwd} with reasonable recommendation times. Particularly at lower values of $k$, the Greedy heuristic offers a balance between speed and reward, making it the preferred approach when recommending a smaller number of courses.

RL approaches demonstrate their strength in efficiency, particularly for recommending longer sequences of courses ($k=4, 5$). PPO, in particular, stands out for maintaining a recommendation quality on par with the Greedy algorithm but with significantly faster execution (one order of magnitude faster than the Greedy approach). This characteristic makes PPO a more suitable option for longer sequence recommendations. Overall, the RL algorithms, DQN and PPO, show potential as the fastest methods, especially in scenarios with larger datasets. When longer sequences of courses are required, RL methods offer a more efficient alternative without compromising the quality of recommendations.

In summary, our results suggest using different approaches to course recommendation based on the sequence length: the Greedy heuristic for shorter sequences where fewer courses are recommended, and RL approaches, particularly PPO, for longer sequences and larger datasets where efficiency becomes paramount.

\section{Limitations}
While our work contributes valuable insights and research directions to the field of job-oriented course recommender systems, it is important to acknowledge several limitations that may impact the generalizability and effectiveness of our recommender system.

\noindent\textbf{Language Restriction.}
Our system currently operates exclusively in English, restricting the applicability of our model in multilingual contexts and excluding non-English job postings and courses.

\noindent\textbf{Dataset Size.}
The datasets used in this study are relatively small, limiting the robustness and scalability of our findings.

\noindent\textbf{Evaluation.}
The lack of expert annotators to validate the course recommendation means our evaluation may not fully capture the system's effectiveness in real-world scenarios. 

\noindent\textbf{Reliance on Heuristics.}
Our approach includes several heuristic methods. While they are necessary for handling data complexities, these heuristics introduce an element of subjectivity and may not always reflect real-world learning and job market dynamics.

\noindent\textbf{Assumption of Skill Acquisition.}
A fundamental assumption in our sequential recommendation approach is that completing a course will automatically result in the acquisition of the associated skills by the learners. This overlooks the variability in individual learning outcomes and the probability of successfully acquiring new skills. A probabilistic approach that estimates the likelihood of skill acquisition would provide a more nuanced and realistic model.

\section{Conclusion}

In this work, we provided the perspective of academic researchers working in collaboration with industry practitioners to develop and deploy a job-market-oriented course recommender system. 
We proposed to rethink course recommender systems to consider the job market, several properties that such systems should satisfy, and research directions that would help develop this field.
We introduced \ses{}, a skill extraction and matching method that efficiently aligns skills extracted from resumes, course content, and job descriptions with the ESCO taxonomy. Utilizing in-context learning and Large Language Models (LLMs), \ses{} is fully unsupervised, enabling 
 generalization to any document type and adaptation to an ever-evolving job market.
Building on this foundation, we developed an unsupervised course recommender system that leverages the matched skills to suggest course sequences aimed at maximizing employment opportunities. 
Our investigation of sequential recommendation strategies — including a greedy heuristic, an exhaustive approach, and two Reinforcement Learning (RL) models — revealed insightful findings. Notably, the RL approaches, particularly Proximal Policy Optimization (PPO), stand out in recommending longer sequences efficiently. They offer a promising solution for larger datasets, balancing recommendation quality with computational efficiency.


\appendix

\section{function design motivation}
\label{sec:appendix:simdesign}
In this section, we motivate our design choices for the user-job similarity function and the user-course relevance function described in section~\ref{subsec:recommendation}. Drawing inspiration from axiomatic Information Retrieval~\cite{DBLP:conf/sigir/FangZ05}, we have designed these similarity functions based on a set of desirable constraints they should fulfill.
\subsection{User-Job Similarity}
\label{subsec:appendix:ljsim}
First, we describe the constraints for the user-job similarity function:\\
\textbf{UJC1:} Assign a higher score to a user who possesses more skills required for a job.\\
\textbf{UJC2:} Assign a higher or equal score to a user with higher proficiency levels in the skills required for a job.\\
\textbf{IJC3:} Assign the maximal score to a user who has all the skills and proficiency levels required for a job.\\

Based on these constraints, we propose the following user-job similarity function, denoted as uj-sim:
\begin{equation}
    \text{uj-sim}(u, j) = \sum\limits_{s \in j} \frac{\text{sim}(\text{sl}_{s, u}, \text{sl}_{s, j})}{|j|}
    \label{eq:ljs}
\end{equation}
\begin{equation}
    \text{sim}(l_i, l_j) = \frac{\min(l_i, l_j)}{l_j}
    \label{eq:simll}
\end{equation}

Here $\text{sl}_{s, j}$ (respectively $\text{sl}_{s, u}$) represents the skill proficiency level for skill $s$ in job $j$ (respectively for user $u$). If a user does not have skill $s$, then $\text{sl}_{s, u} = 0$. The use of $\min$ in the skill-skill similarity function (equation~\ref{eq:simll}), ensures that users matching all requirements will achieve maximum similarity, satisfying constraint \textbf{UJC3}.

\subsection{User-Course Relevance}
\label{subsec:appendix:lcsim}

Next, we define the constraints for the user-course relevance:\\
\textbf{UCC1:} Assign a higher score to a user who possesses more skills required for a course.\\
\textbf{UCC2:} Assign a higher or equal score to a user with higher proficiency levels in the skills required for a course.\\
\textbf{UCC3:} Assign the minimal score to a user who already possesses all the skills and proficiency levels provided by a course.\\
\textbf{UCC3} is based on the rationale that if a user already knows everything taught by a course, they will gain no new skills making the course irrelevant.

We propose the following user-course relevance function, uc-rel:

\begin{equation}
    \text{uc-rel}(u, c) = \text{uc}_{\text{r}}(u, c_r) \cdot (1 - \text{uc}_{\text{p}}(u, c_p))
    \label{eq:uc}
\end{equation}

\begin{equation}
    \text{uc}_{\text{r}}(u, c_r) = \sum\limits_{s \in c_r} \frac{\text{sim}(\text{sl}_{s, u}, \text{sl}_{s, c_r})}{|c_r|}
    \label{eq:lcr}
\end{equation}

\begin{equation}
    \text{uc}_{\text{p}}(u, c_p) = \sum\limits_{s \in c_p} \frac{\text{sim}(\text{sl}_{s, u}, \text{sl}_{s, c_p})}{|c_p|}
    \label{eq:lcp}
\end{equation}

In these equations, \(\text{sl}_{s, c_r}\) (respectively \(\text{sl}_{s, c_p}\)) indicates the required (respectively provided) proficiency level for skill \(s\) in course \(c\). Because $\text{uc}_{\text{p}}(u, c_p)$ is bounded between 0 and 1 (from equation~\ref{eq:simll}) and returns 1 if and only if user $u$ already possesses all the skills provided by the course, the overall relevance $\text{uc-rel}(u, c)$ will be equal to 0 (the lowest possible value) ensuring constraint \textbf{UCC3}.

\bibliographystyle{ACM-Reference-Format}
\bibliography{sample-base}

\end{document}